\documentclass[iop,apj]{emulateapj}   

\usepackage{graphicx}        
\usepackage{amssymb}         
\usepackage[usenames,dvipsnames]{color}          
\definecolor{darkblue}{rgb}{0,0,0.5}
\usepackage{epstopdf}

\DeclareFontFamily{OT1}{pzc}{}
\DeclareFontShape{OT1}{pzc}{m}{it}%
             {<-> s * [1.1500] pzcmi7t}{}
\DeclareMathAlphabet{\mathscr}{OT1}{pzc}%
                                 {m}{it}

\usepackage{amsmath}

\newcommand{\fract}[2]{\leavevmode\kern.1em
          \raise.5ex\hbox{\the\scriptfont0 #1}\kern-.1em
    \raise.15ex\hbox{\the\scriptfont0 /}\kern-.08em\lower.25ex\hbox{\the\scriptfont0 #2}}

\newcommand{\half}{{\textstyle\frac{1}{2}}}

\newcommand{\e}{\hat{\mathbf{e}}}
\newcommand{\g}{\mathbf{g}}
\renewcommand{\k}{\mathbf{k}}

\newcommand{\bj}{\mathbf{j}}

\newcommand{\boldv}{{\mathbf{v}}}

\newcommand{\B}{{\mathbf{B}}}
\newcommand{\E}{{\mathbf{E}}}
\newcommand{\F}{{\mathbf{F}}}

\newcommand{\D}{\mbox{\boldmath$\mathsf{D}$}}

\newcommand{\vdot}{{\boldsymbol{\cdot}}}
\newcommand{\vcross}{{\boldsymbol{\times}}}
\newcommand{\grad}{\mbox{\boldmath$\nabla$}}
\newcommand{\bxi}{\mbox{\boldmath$\xi$}}

\newcommand{\thth}{\hspace{1.5pt}}
\newcommand{\curl}{\grad\vcross}
\newcommand{\Curl}{\grad\vcross\thth}

\newcommand\Div{\grad\vdot\thth}

\newcommand{\ri}{{i}}

\renewcommand{\leq}{\leqslant}  \renewcommand{\le}{\leqslant}

\allowdisplaybreaks[1]

\begin{document}


\title{Fast-to-Alfv\'en Mode Conversion Mediated by Hall Current\\
 I. Cold Plasma Model}

\author{Paul S.~Cally\altaffilmark{1}}
\affil{School of Mathematical Sciences and Monash Centre for Astrophysics,\\ Monash University, Clayton, Victoria 3800, Australia}
  \email{paul.cally@monash.edu}
  
  \and
  
\author{Elena Khomenko}
\affil{Instituto de Astrof\'isica de Canarias, 38205 La Laguna, Tenerife, Spain and \\
Departamento de Astrof\'{\i}sica, Universidad de La Laguna, 38205, La Laguna, Tenerife, Spain}
\email{khomenko@iac.es}

\altaffiltext{1}{Jesus Serra Foundation Fellow at the Instituto de Astrof\'isica de Canarias, Tenerife, Spain, May--June 2015}



\begin{abstract}\noindent

The photospheric temperature minimum in the Sun and solar-like stars is very weakly ionized, with ionization fraction $f$ as low as $10^{-4}$. In galactic star forming regions, $f$ can be $10^{-10}$ or lower. Under these circumstances, the Hall current can couple low frequency Alfv\'en and magneto\-acoustic waves via the dimensionless Hall parameter $\epsilon=\omega/\Omega_\text{i}f$, where $\omega$ is the wave frequency and $\Omega_\text{i}$ is the mean ion gyrofrequency. This is analysed in the context of a cold (zero-$\beta$) plasma, and in less detail for a warm plasma. It is found that Hall coupling preferentially occurs where the wave vector is nearly field-aligned. In these circumstances, Hall coupling in theory produces a continual oscillation between fast and Alfv\'en modes as the wave passes through the weakly ionized region. {\color{black} At low frequencies (mHz), characteristic of solar and stellar normal modes, $\epsilon$ is probably too small for more than a fraction of one oscillation to occur. On the other hand, the effect may be significant at the far higher frequencies (Hz) associated with magnetic reconnection events. In another context,} characteristic parameters for star forming gas clouds suggest that $\mathcal{O}(1)$ or more full oscillations may occur in one cloud crossing. This mechanism is not expected to be effective in sunspots, due to their high ion gyrofrequencies and Alfv\'en speeds, since the net effect depends inversely on both and therefore inverse quadratically on field strength.
\end{abstract}



\section{Introduction}
Alfv\'en waves \citep{Alf42aa} are the archetypal magneto\-hydro\-dynamic wave. They are due to magnetic tension only, whereas the fast and slow magneto\-acoustic waves involve gas and magnetic pressure as well. In gravitationally stratified atmospheres such as those of the Sun and stars, these wave types do not necessarily retain their identities over many scale heights.

For example, the fast and slow magneto\-acoustic waves may mode convert in stellar atmospheres at the Alfv\'en-acoustic equipartition level, $a=c$ where $a$ is the Alfv\'en speed and $c$ is the sound speed \citep{Cal06aa,SchCal06aa}. Fast-to-Alfv\'en conversion occurs near the fast wave reflection height caused by increasing Alfv\'en speed with height in gravitationally stratified atmospheres \citep{CalHan11aa,KhoCal11aa,HanCal12aa,KhoCal12aa,Fel12aa}, provided the wave vector is not in the same plane as gravity and the magnetic field.

Fast-to-Alfv\'en conversion is potentially important in supplying wave energy and heating to solar and stellar coronae and winds. Whereas fast waves typically reflect totally in the high chromosphere or from the Transition Region (TR), Alfv\'en waves are more able to penetrate to the outer atmosphere. \citet{HanCal12aa,HanCal14ac} found that Alfv\'en waves generated by mode conversion in the upper solar chromosphere are far more able to penetrate the TR than are those generated at the photosphere. ``Alfv\'enic'' waves have recently been observed with sufficient amplitude to heat the Sun's quiet-region corona and power the fast solar wind \citep{McIde-Car11aa}. The presumably varied origins of Alfv\'en waves are therefore of great importance in understanding the fundamental nature of the outer atmospheres of solar-type stars and their winds.

Previous modelling of mode conversion between fast and Alfv\'en waves has been based on ideal magneto\-hydro\-dynamics (ideal MHD). The temperature minimum region of the Sun and solar-like stars though is so weakly ionized that the multi-component nature of the plasma should be taken into account \citep{KhoColDia14aa}. We focus on the Hall current, which is likely the dominant non-MHD effect around the quiet Sun temperature minimum \citep[see][Fig.~6a]{KhoColDia14aa}. The relative orders of the Ohmic, ambipolar, and Hall terms are discussed in some generality by \citet{PanWar08aa}. 


At first sight, Hall current has the potential to couple Alfv\'en and magneto\-acoustic waves by ``mixing'' their polarizations in some way (the velocity polarizations of the Alfv\'en and two magneto-acoustic waves are mutually orthogonal in ideal MHD). This is based on the analysis and numerical simulations of \citet{CheCam12aa}, who argue that Hall current precesses the polarization of field-aligned Alfv\'en waves in a uniform plasma. This interpretation is not strictly true; we shall see that the Hall term actually causes waves to oscillate between magneto\-acoustic and Alfv\'en states, with the ``precession'' being a beating between the nearly degenerate magneto\-acoustic and Alfv\'en modes. 

\textcolor{black}{Fast/Alfv\'en coupling by the Hall effect has also been noted in the ionospheric literature \citep{KatTam56aa,WatLysSci13aa}, but the precessional nature is not brought out.}

We first examine the Hall-coupling phenomenon in a simple exponentially stratified cold (zero-$\beta$) plasma, in direct comparison to the ideal MHD analysis of \cite{CalHan11aa}. More realistic warm plasmas are addressed briefly in Section \ref{hot}, and will be explored further in Paper II.

\par\penalty-1000
\section{Mathematical Formulation}
\subsection{Wave Equations in a Stratified Atmosphere}
Following \citet{CalHan11aa}, we consider a cold MHD plasma with uniform magnetic field $\B_0=B_0(\cos\theta,0,\sin\theta)$. In the cold plasma (also called the zero-$\beta$) approximation, the sound speed is neglected in the perturbation equations compared to the Alfv\'en speed. This freezes out slow MHD waves, leaving only fast and Alfv\'en waves. The density stratification due to gravity (assumed to act in the negative $x$-direction) remains though, and produces an Alfv\'en speed $a$ that increases (we shall assume) exponentially with $x$: $a(x)=a_0\exp[x/2h]$, where $h$ is the density scale height. See Appendix \ref{app:der} for details.

\begin{figure}[htb]
\begin{center}
\includegraphics[width=.8\hsize]{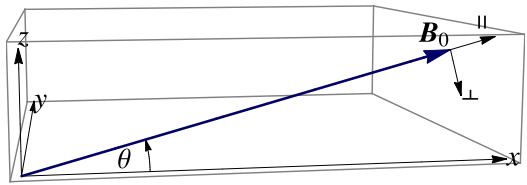}
\caption{Coordinate systems used to model the oscillations.  Density and Alfv\'en speed vary in the $x$-direction only. $\B_0$, $\e_\parallel$, and $\e_\perp$ are all in the $x$-$z$ plane, so $(\e_\perp,\e_y,\e_\parallel)$ form a right-handed orthogonal coordinate system.}
\label{fig:coords}
\end{center}
\end{figure}

Hall current contributes an additional term to the electric field, which becomes
\begin{equation}
\E=-\boldv\vcross\B+\frac{\bj\vcross\B}{e n_\text{e}},
\end{equation}
where $\boldv$ is the mass-weighted combination of electron and ion fluid velocities, $e$ is the elementary charge and $n_\text{e}$ is the electron number density \citep[see][p.~173]{GoeKepPoe10aa}. The current density $\bj=\mu_0^{-1}\Curl\B$ (neglecting the displacement current as usual) and the Faraday equation $\partial\B/\partial t =  -\Curl\E$ may be combined with the momentum equation $\rho\thth D\boldv/Dt=\bj\vcross\B$ to complete the description of the system. It is assumed that the plasma is collisionally dominated, so that the inertia of the neutrals plays a full role in the oscillations. Consequently, the full mass density $\rho$ appears in the Alfv\'en speed $a=B/\sqrt{\mu_0\rho}$, and not just the ion density.

The linearized Hall-MHD equations may be combined into a single vector equation in the plasma displacement $\bxi$ (see Appendix \ref{app:der}). Fourier analysing in time, $\bxi(x,y,z,t)=\bxi(x,y,z)\exp(-\ri\omega t)$, and introducing the ``Hall parameter''
\begin{equation}
\epsilon=\frac{\omega}{f\,\Omega_\text{i}},  \label{epsilon}
\end{equation}
where $f$ is the ionization fraction and $\Omega_\text{i}$ the mean ion gyrofrequency, these equations may be combined to yield
\begin{equation}
\left(\partial_\parallel^2+\frac{\omega^2}{a^2}\right)\bxi  =-\grad_{\!\text p}\chi
+i\left[\grad\tilde\chi\vcross\e_\parallel-\nabla^2\left(\tilde\bxi\vcross\e_\parallel\right)\right]
,                               \label{basiceqn}
\end{equation}
where  $\chi=\Div\bxi$ is the dilatation. The Hall-parameter-scaled displacement $\tilde\bxi=\epsilon\,\bxi$ is also introduced, with $\tilde\chi=\Div\tilde\bxi$. Furthermore, $\e_\parallel=\hat\B_0$ is the unit vector in the direction of the magnetic field, $\partial_\parallel=\hat\B_0\vdot\grad$ is the field-aligned directional derivative, and $\grad_{\!\text p}=\grad-\hat\B_0\,\partial_\parallel$ is the complementary perpendicular component of the gradient. The coordinates used are illustrated in Figure \ref{fig:coords}.

The ionization fraction is $f=m_\text{i}n_\text{i}/\rho$, with $m_\text{i}$  and $n_\text{i}$ the mean ion mass and total number density respectively. The mass of the electron is neglected relative to that of the ion. Charge neutrality $n_\text{e}=Z n_\text{i}$ is assumed, where $Z$ is the ion mean charge state. The ion gyrofrequency is denoted by $\Omega_\text{i}=Z e B_0/m_\text{i}$. Equation (\ref{basiceqn}) generalizes Equation (1) of \citet{CalHan11aa} by the addition of a term proportional to the ratio of the wave frequency to the ion gyrofrquency and inversely to the ionization fraction. For waves of helioseismic interest (frequencies of a few mHz), $\omega/\Omega_\text{i}$ is very small (the proton gyrofrequency is $15.2\,B_0$ MHz for example, where $B_0$ is measured in Tesla), suggesting that very small ionization fractions and low field strengths are required for there to be any significant effect. This is discussed further in Section \ref{min}.


The plasma displacement is conveniently expressed in either the $(x,y,z)$ Cartesian coordinate system
$\bxi(x)=\xi_x\e_x+\xi_y\e_y+\xi_z\e_z$, or in the magnetic flux system defined by the parallel direction (denoted by `$\scriptstyle\parallel$'), the direction $(\sin\theta,0,-\cos\theta)$ perpendicular to $\B_0$ but in the plane containing the field lines and the direction of stratification (denoted by `$\scriptstyle\perp$'), and the $y$ direction perpendicular to both: $\bxi(x)= \xi_\perp \e_\perp+\xi_y\e_y$. There is no component of displacement in the parallel direction as there is no restoring force that acts in that direction, having lost the gas pressure perturbation in the cold plasma approximation. We also use the subscript `p' to denote the plane perpendicular to $\B_0$, \emph{i.e.}, spanned by the unit basis vectors $\e_\perp$ and $\e_y$. Decomposing Equation (\ref{basiceqn}) into the two components, and Fourier analysing in both time and the homogeneous spatial dimensions $\bxi(x,y,z,t)=\bxi(x)\exp[\ri(k_yy+k_zz-\omega t)]$, 
\begin{subequations}\label{parxi}
\begin{multline} 
\left(\partial_\parallel^2+\partial_\perp^2+\frac{\omega^2}{a^2}\right)\xi_\perp
 = -\ri\, k_y \partial_\perp\xi_y \\
 +i
 \left[
 i\,k_y\partial_\perp(\epsilon\,\xi_\perp)-(\partial_\parallel^2+\partial_\perp^2)(\epsilon\,\xi_y)
 \right] \label{parxiA}
\end{multline}
and
\begin{multline}
\left(\partial_\parallel^2+\frac{\omega^2}{a^2}-k_y^2\right)\xi_y = -\ri\, k_y \partial_\perp\xi_\perp\\
 +i
 \left[
 (\partial_\parallel^2-k_y^2)(\epsilon\,\xi_\perp)-i\,k_y\partial_\perp(\epsilon\,\xi_y)
 \right],   \label{parxiB}                
\end{multline} 
\end{subequations}
where $\partial_\perp=\e_\perp\!\vdot\thth\grad$, and of course $\partial_\parallel^2+\partial_\perp^2=\partial_x^2-k_z^2$ is just the 2D Laplacian in the $(\parallel,\perp)$ or $(x,z)$ plane.

The derivatives $\partial_\parallel=\cos\theta\,\partial_x+\sin\theta\,\partial_z=\cos\theta\,\partial_x+i\,k_z\sin\theta$ and $\partial_\perp=\sin\theta\,\partial_x-\cos\theta\,\partial_z=\sin\theta\,\partial_x-i\,k_z\cos\theta$ cannot be completely written in terms of wavenumbers when the background depends on $x$, except in the WKB weakly inhomogeneous approximation \citep[see][for example]{Whi74aa,BenOrs78aa}, which we will use  for expository purposes in Sections \ref{local} and \ref{osc}, and in part in Section \ref{hot}. Otherwise though, the exact equations are retained. 

As emphasised by \citet{CalHan11aa}, cross-field wave propagation, $k_y\ne0$, couples the fast and Alfv\'en waves, characterized respectively by $\xi_\perp$ and $\xi_y$ when $k_y$ is small. Equations (\ref{parxi}) indicate that the Hall terms do this too, even when $k_y=0$.

Strictly, pure Alfv\'en waves exist only in the two-dimensional (2D) case $k_y=0$, as only then is there a direction ($\e_y$) perpendicular to both the magnetic field and wave propagation direction in which the background medium is invariant. However, even with $k_y\ne0$, there are waves that are ``asymptotically Alfv\'en'' as $x\to\infty$, as shown by Frobenius expansion \citep[see the Appendix to][]{CalHan11aa}.

Assuming an exponentially decreasing density of scale height $h$, we may set the dimensionless quantity $\omega^2h^2/a^2=e^{-x/h}$, thereby arbitrarily fixing the zero point of $x$. Note that this zero point is dependent on frequency as well as the background medium properties. The classical reflection point of the fast wave is at $\omega=a\,k$, where $k^2=k_y^2+k_z^2$, or equivalently $\omega h/a=\kappa$, where $\kappa=kH$ is the dimensionless wave number transverse to $x$. The significance of $x=0$ then is that it is the point at which fast waves of transverse wavenumber $\kappa=1$ reflect. We shall mostly be concerned with $\kappa\ll1$ for which transverse wavelength is much larger than the density scale height.

\goodbreak
\subsection{Local Analysis}  \label{local}
{\color{black}The local analysis of a cold plasma with Hall current is discussed in some detail by \citet{DamWriMcK09aa}, Section III.B, for a fully ionized plasma. However, we re-assess the equations here with a particular focus on a ``precession'' or oscillation between the fast and Alfv\'en waves that does not seem to have been fully appreciated up till now.}

Neglecting all spatial variation in the background, or equivalently assuming all wavelengths are small compared to the density scale height, we may identify $\partial_\parallel=i\,k_\parallel$ and $\partial_\perp=i\,k_\perp$. Then Equations (\ref{parxiA}) and (\ref{parxiB}) may be reduced to algebraic matrix form $\D\bxi=\mathbf{0}$ where
\begin{multline}
\D=\\
\begin{pmatrix}
\frac{\omega^2}{a^2}-k_\parallel^2-k_\perp^2+i\,\epsilon\,k_yk_\perp & -k_yk_\perp-i\,\epsilon(k_\parallel^2+k_\perp^2)\\[4pt]
-k_yk_\perp+i\,\epsilon(k_\parallel^2+k_y^2) & \frac{\omega^2}{a^2}-k_\parallel^2-k_y^2-i\,\epsilon\,k_yk_\perp,
\end{pmatrix}
\end{multline}
$\bxi=(\xi_\perp,\xi_y)^T$. The determinant of the coefficient matrix specifies the dispersion relation,
\begin{equation}
\mathscr{D}=\det\D=(\omega^2-a^2k_\parallel^2)(\omega^2-a^2 k^2) - \epsilon^2a^4k_\parallel^2 k^2=0,
\end{equation}
where $k^2=|\k|^2=k_\parallel^2+k_\perp^2+k_y^2$, {\color{black} in accord with Equation (17) of \citet{DamWriMcK09aa} (for $f=1$)}. The Hall term couples the otherwise disjoint Alfv\'en wave, $\omega^2=a^2k_\parallel^2$, and fast wave, $\omega^2=a^2k^2$. The eigenvectors, specifying the wave polarizations, are
\begin{multline}
\bxi_\text{fast},\,\bxi_\text{A} =\\
\left[
k_\perp^2-k_y^2-2i\,k_\perp k_y\epsilon \pm
\sqrt{(k_\perp^2+k_y^2)^2+4k_\parallel^2k^2\epsilon^2}\,
\right]\e_\perp\\
+2\left[k_\perp k_y-i(k_\parallel^2+k_y^2)\epsilon\right]\e_y
\end{multline}
for the fast ($+$ sign) and Alfv\'en ($-$ sign) modes respectively. Due to the imaginary terms proportional to $\epsilon$, these describe elliptical polarization for $\epsilon\ne0$.

In ideal MHD, the case $k_\perp=k_y=0$, $k=k_\parallel$, is degenerate and the eigenvectors are arbitrary in the normal plane. The Hall term however splits the degeneracy, with eigenvalues satisfying
\begin{equation}
(\omega^2-a^2k^2)^2=\epsilon^2a^4k^4, \quad\text{\emph{i.e.},}\quad\omega^2 = a^2 k^2 \pm \epsilon\, a^2 k^2 , \label{parDR}
\end{equation}
and eigenvectors $\pm i\,\e_\perp+\e_y$. This dispersion relation accords with Equation (14.128) of \citet{GoeKepPoe10aa} (for $f=1$). The small $\mathcal{O}(\epsilon)$ difference between the eigenfrequencies of the Alfv\'en and fast modes results in precession of the polarization of the combined ``Alfv\'en wave" $\omega^2=a^2k^2$, with precession frequency $\half(\sqrt{1+\epsilon}-\sqrt{1-\epsilon})ak=\half\epsilon ak+\mathcal{O}(\epsilon^3)$, where
 $\epsilon=ak/\Omega_if$ to leading order. This is the phenomenon of beating, and was noted by \citet{CheCam12aa}.
It does not occur far away from the $k_\perp=k_y=0$ degeneracy, or at least, the fast and Alfv\'en waves are more distinct there, and so would not in combination be perceived as a single precessing mode.

The analysis of \citet{CheCam12aa}, Section 3.1, returns dispersion relation $\omega^2=a^2k^2+\sigma^2$ (their Equation (16)) where $\sigma=(\epsilon/2)ak$ in our notation, ignoring the $\omega^2=a^2k^2-\sigma^2$ solution. This is at odds with our value from Equation (\ref{parDR}) above, $\sqrt{\epsilon}\,ak$, and Equation (14.128) of \citet{GoeKepPoe10aa} (for $f=1$). However, they then state that the precession frequency is (their) $\sigma$, \emph{i.e.}, $(\epsilon/2)ak$, which is the correct formula to $\mathcal{O}(\epsilon^2)$. A more correct statement of their formulation would have been $\omega=ak+\sigma$, so that $\omega^2=a^2k^2+2\sigma a k+\mathcal{O}(\sigma^2)$, or in other words $\omega^2=a^2k^2+\epsilon\, a^2k^2$ to $\mathcal{O}(\epsilon)$. Their analysis though was hampered by the assumption of an original ansatz that did not recognize the implicit coupling with the fast wave that makes the dispersion relation quartic in $\omega$. The polarizations of single normal modes do not precess; they are just the unique eigenvectors.


Nevertheless, as depth increases in near-vertical field, with $k_y$ and $k_z$ fixed, the longitudinal Alfv\'en wavenumber $k_\parallel\approx k_x\approx\omega/a$ increases exponentially and the degenerate case is approached. We shall see that this is indeed the circumstance in which Hall-induced mode conversion occurs most strongly, mediated by precession of the joint polarization of the fast and Alfv\'en modes. This is discussed further in Section \ref{osc}. 

\par\penalty-2000
\subsection{Oscillatory Behaviour with $\epsilon$ in Unstratified or Weakly Stratified Atmospheres} \label{osc}

If we neglect the variation of $a$ and $\epsilon$ with $x$, the basic equation (\ref{basiceqn}) may be recast in terms of $\chi=\Div\bxi$ and $\zeta=\e_\parallel\vdot\Curl\bxi$, which perfectly characterize the fast and Alfv\'en waves respectively at all wave orientations:
\begin{subequations}\label{chizeta}
\begin{align}
\left(\nabla^2+\frac{\omega^2}{a^2}\right)\chi &=-i\,\epsilon\,\nabla^2\zeta\label{chizetaA}\\
\left(\partial_\parallel^2+\frac{\omega^2}{a^2}\right)\zeta &=i\,\epsilon\,\partial_\parallel^2\chi.\label{chizetaB}
\end{align}
\end{subequations}
Only the Hall term couples the two modes now. 

Further neglecting the cross-field derivatives, $\partial_\perp=\partial_y=0$, the dispersion relation is transparently just as set out in Equation (\ref{parDR}). The solution for $\chi$ and $\zeta$ contains linear combinations of trigonometric terms with rapid (Alfv\'enic) oscillations and slowly varying sinusoidal amplitudes.

For comparison with later numerical solutions, it is now convenient to
solve for $k$ with fixed wave frequency $\omega$ yielding four roots, $k = \pm\omega/(a\sqrt{1\pm\epsilon})$,
where the two $\pm$ signs are independent. The corresponding modes are therefore
\begin{equation}
\exp\left[ \pm i \frac{\omega\,s/a}{\sqrt{1\pm\epsilon}}\right].  \label{kmodes}
\end{equation}
These combine to produce an Alfv\'enic wavenumber
\begin{equation}
k_\text{Alf}=\frac{\omega}{2a}\left(\frac{1}{\sqrt{1-\epsilon}}+\frac{1}{\sqrt{1+\epsilon}}\right)
 = \frac{\omega}{a}+\mathcal{O}(\epsilon^2)
\end{equation}
and an envelope wavenumber
\begin{equation}
k_\text{env}=\frac{\omega}{2a}\left(\frac{1}{\sqrt{1-\epsilon}}-\frac{1}{\sqrt{1+\epsilon}}\right)
=\frac{\epsilon\,\omega}{2a}+\mathcal{O}(\epsilon^3).
\end{equation}
The envelope is produced by the beating of the near-degenerate modes and describes an oscillatory transfer of energy between the compressive fast wave ($\chi$) and incompressive Alfv\'en mode ($\zeta$) on their journey through the Hall window. This will be confirmed numerically and generalized in propagation direction in Section \ref{2D3D}. The spatial periodicity $4\pi a/\epsilon\omega$ (for small $\epsilon$) corresponds to a temporal periodicity $4\pi/\epsilon\omega$, \emph{i.e.}, circular frequency $\epsilon a k/2$, which is just the ``precession'' frequency identified in Section \ref{local}.



\subsection{Overview}
Elementary analysis of the governing wave equations has already told us a lot. Both out-of-the-plane wave orientation $k_y$ and Hall current $\epsilon$ couple fast and Alfv\'en waves, but in very different fashions. The former requires Alfv\'en speed stratification and operates locally near the fast wave reflection point. The latter applies everywhere that $\epsilon\ne0$, even in an unstratified plasma.

For a fixed Hall-effective window of thickness $L$, we must expect the fast-to-Alfv\'en conversion coefficient $\mathscr{A}^+$ (see Section \ref{pertgenform}) to display a $2\pi a/\omega L$ periodicity in $\epsilon$ when the wavevector is field-aligned. Account has been taken of the quadratic dependence of $\mathscr{A}^+$ on $\zeta$, making the periodicity $2\pi a/\omega L$ rather than $4\pi a/\omega L$. With increasing $L$ or decreasing Alfv\'en speed $a$, the conversion coefficient oscillates ever more rapidly with $\epsilon$. 
In a slowly varying atmosphere, the number of oscillations in passing from $x_1$ to $x_2$ would be 
\begin{equation}
N_\text{osc}=\int_{x_1}^{x_2}\frac{\epsilon\thth\omega}{2\pi a}\, dx,  \label{n}
\end{equation}
which shall be termed the ``oscillation number''. A half-integer value corresponds to total conversion, whilst a full integer yields zero net conversion. In practice, the Hall parameter $\epsilon$ and Hall window thickness $L$ may be small enough or the Alfv\'en speed large enough that this periodic behaviour is never seen for low-frequency waves. 

For fixed frequency, the inverse quadratic dependence of the oscillation number on magnetic field strength is important (one factor of $B_0^{-1}$ through the ion gyrofrequency in $\epsilon$, and the other through the Alfv\'en speed in the denominator). Ultimately, this means that Hall conversion can never be effective for mHz-frequency waves in sunspots (see Section \ref{min}).

We now turn to arbitrary magnetic field and wavevector orientations. In Section \ref{pert} we address the two dimensional (2D) case $k_y=0$ using a perturbation analysis. Both 2D and 3D cases are examined numerically in Section \ref{2D3D}.

\begin{figure*}[tbhp]
\begin{center}
\includegraphics[width=\hsize]{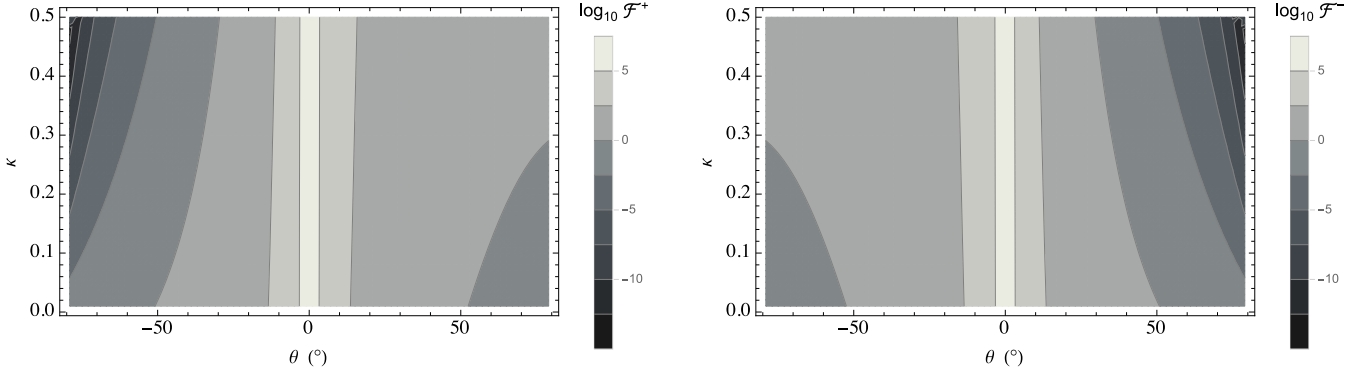}
\caption{Contour plots of $\mathscr{F}^+$ (left) and $\mathscr{F}^-$ (right) against magnetic field inclination from the vertical $\theta$ and dimensionless horizontal wavenumber $\kappa=k_zh$. Note the logarithmic scale. Both functions are singular on the line $\theta=0$.}
\label{fig:FPM}
\end{center}
\end{figure*}

\goodbreak
\section{Hall Conversion in the 2D Case through Perturbation Analysis}\label{pert}\nobreak
\subsection{General Formulation}\label{pertgenform}\nobreak
In the 2D case $k_y=0$, the Hall terms proportional to $\omega/\Omega_\text{i}$ still couple $\xi_\perp$ and $\xi_y$. To understand the nature of the coupling, it is useful to perform a perturbation analysis, expanding to first order in $\epsilon=\omega/\Omega_\text{i}f$. Typically, $\epsilon\ll1$  for waves of interest in the low solar atmosphere. At this stage, it is convenient to scale all lengths by $h$, or equivalently set $h=1$.

We consider a (zeroth order) fast wave defined by
\begin{equation}
\left(\partial_\parallel^2+\partial_\perp^2+\frac{\omega^2}{a^2}\right)\xi_{\perp0}=\left(\partial_x^2-k_z^2+\frac{\omega^2}{a^2}\right)\xi_{\perp0}=0,
\end{equation}
with $\xi_{y0}=0$. With $\omega^2/a^2=e^{-x}$, the appropriate solution is a Bessel function of the first kind,
\begin{equation}
\xi_{\perp0}=J_{2\kappa}\left(2 e^{-x/2}\right),
\end{equation}
representing the reflecting fast wave that is evanescent as $x\to\infty$. Here $\kappa=k_z h$ is the dimensionless $z$-wavenumber.

The Alfv\'en wave driven through Hall coupling by this solution satisfies
\begin{equation}
\left(\partial_\parallel^2+\frac{\omega^2}{a^2}\right)\xi_{y1}=i\,\epsilon\,R_H(x),  \label{y1}
\end{equation}
where 
\begin{equation}
R_H =  f\,\partial_\parallel^2 \left(f^{-1}{\xi_{\perp0}} \right) .
   \label{RH}
\end{equation}

Equation (\ref{y1}) may be solved using the method of variation of parameters, or equivalently by constructing a Green's function. The solutions of the homogeneous equations are most conveniently expressed as Hankel functions $e^{-i\kappa x\tan\theta}H_0^{(1)}(2 e^{-x/2}\sec\theta)$ and $e^{-i\kappa x\tan\theta}H_0^{(2)}(2 e^{-x/2}\sec\theta)$, representing respectively the leftward (downward) and rightward (upward) propagating Alfv\'en waves. Their Wronskian is $W=2\,i\,\pi^{-1}\exp[-2i \kappa\,x\tan\theta]$. The formal driven inhomogeneous solution is then
\begin{multline}
\xi_{y1} = \epsilon\,\frac{\pi}{2}\, e^{i\kappa\,x \tan\theta} \sec^2\theta
\left[ A_1(x) H_0^{(1)}(2 e^{-x/2}\sec\theta)\right. \\ \left.+ A_2(x) H_0^{(2)}(2 e^{-x/2}\sec\theta)\right],
\end{multline}
where
\begin{align}
A_1(x)&=\int_x^\infty e^{i\,\kappa\,X\tan\theta} H_0^{(2)}(2 e^{-X/2}\sec\theta)\,R_H(X)\,dX\nonumber\\[0pt]
 &= \int_0^{e^{-x}} s^{-i\,\kappa\tan\theta-1}H_0^{(2)}(2 \sqrt{s}\sec\theta)\,R_H(-\ln s)\,ds \nonumber\\[0pt]
 &=  \int_{e^{-x}}^\infty a_1(s)\,ds,\\
\noalign{\noindent and}\nonumber\\
A_2(x)&=\int_{-\infty}^x e^{i\,\kappa\,X\tan\theta} H_0^{(1)}(2 e^{-X/2}\sec\theta)\,R_H(X)\,dX\nonumber\\[0pt]
 &= \int_{e^{-x}}^\infty s^{-i\,\kappa\tan\theta-1}H_0^{(1)}(2 \sqrt{s}\sec\theta)\,R_H(-\ln s)\,ds\nonumber\\[0pt]
& =  \int_{e^{-x}}^\infty a_2(s)\,ds,
\end{align}
assuming the integrals exist. The change of variable $s=e^{-x}\in(0,\infty)$ has been introduced.

Conversion to the upgoing Alfv\'en wave is determined by the wave-energy conversion coefficient \citep[see][]{CalHan11aa}
\begin{equation}
\mathscr{A}^+=\epsilon^2\pi^2\sec^2\theta \left| A_2(\infty)\right|^2=\epsilon^2\mathcal{F}^+,  \label{A1}
\end{equation}
defining $\mathcal{F}^+$. Conversely, conversion to the downgoing Alfv\'en wave has coefficient
\begin{equation}
\mathscr{A}^-=\epsilon^2\pi^2\sec^2\theta \left| A_1(-\infty)\right|^2=\epsilon^2\mathcal{F}^-.  \label{A2}
\end{equation}
$\mathscr{A}^\pm=0$ indicates no conversion, and 1 means total conversion. The perturbation solution breaks down before either reach 1 as $\epsilon$ increases \citep[see Fig.~9 of][]{CalHan11aa}.

\begin{figure}[tbhp]
\begin{center}
\includegraphics[width=\hsize]{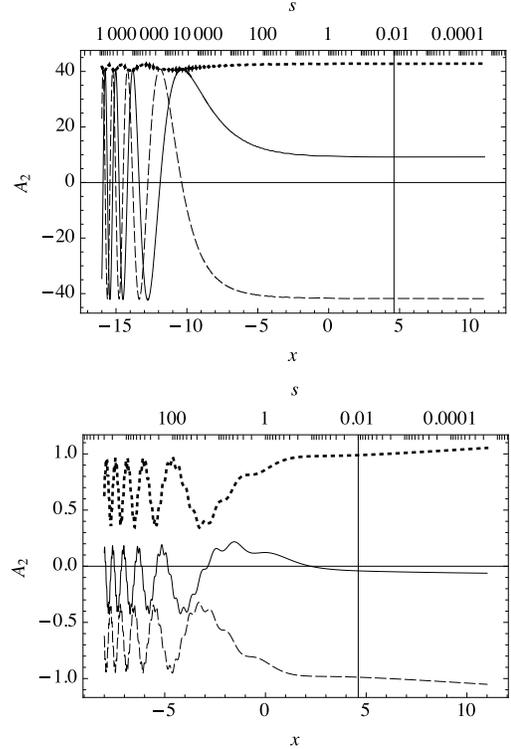}
\caption{Interaction integral $A_2$ in 2D ($k_y=0$) for $\kappa=0.1$ with $\theta=5^\circ$ (top) and $40^\circ$ (bottom). The real and imaginary parts are represented by full and dashed curves respectively, and the absolute value $|A_2|$ is shown by the dotted curves. The vertical line indicates the fast wave reflection point.}
\label{fig:A2}
\end{center}
\end{figure}

\begin{figure}[tbhp]
\begin{center}
\includegraphics[width=0.8\hsize]{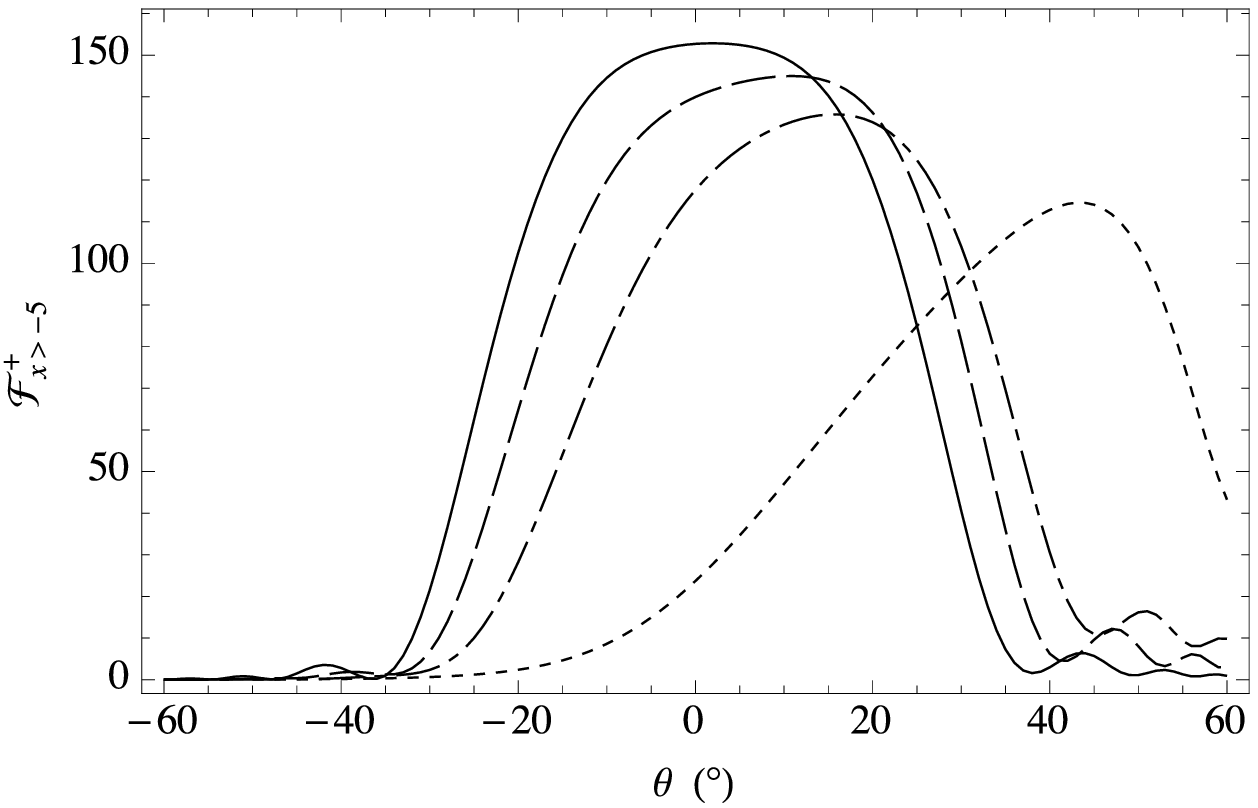}\vspace{8pt}
\includegraphics[width=0.8\hsize]{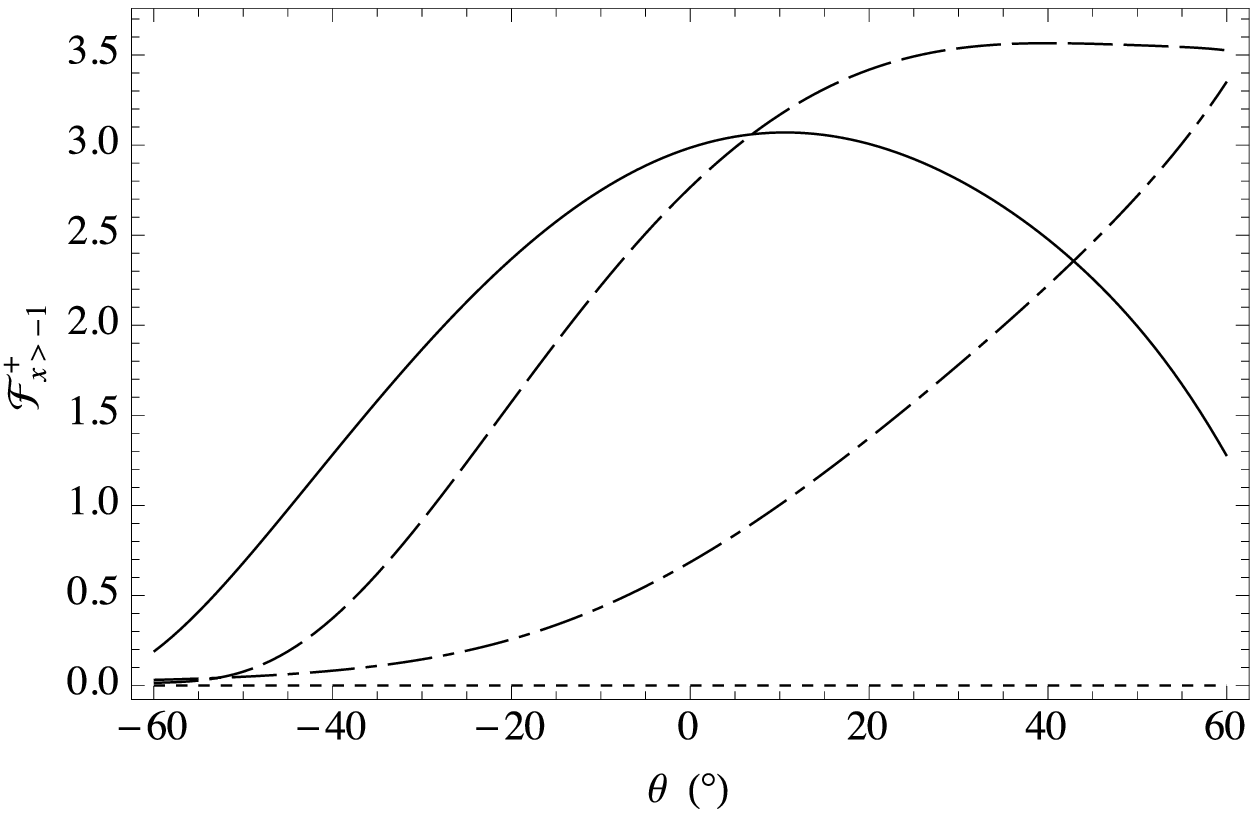}
\caption{$\mathscr{F}^+$ integrated over $x>-5$ only (top panel) and $x>-1$ only (bottom panel) as a function of field inclination $\theta$ for four values of $\kappa$: 0.1 (full); 0.5 (long dashed); 1 (chained); and 4 (short dashed).}
\label{fig:FplusP5}
\end{center}
\end{figure}

\subsection{Uniform Ionization Fraction}
To gain an understanding of the Hall coupling process, we first explore the case of uniform ionization fraction $f$ in a partially ionized gas, for which $\epsilon$ is uniform. Then
\begin{equation}\label{RHU}
\begin{split}
&R_H(-\ln s) =  \partial_\parallel^2 \xi_{\perp0} \\
&\quad = s^{\kappa+1} \left[\, _0\tilde{F}_1\left(;2
   \kappa +1;-s\right) \left(
  \frac{ \kappa^2 }{s}e^{2 i \theta }-\cos ^2\theta\right)\right.\\
  &\qquad\left. - \frac{i\,\kappa\sin 2\theta}{s}
       \,
   _0\tilde{F}_1\left(;2 \kappa
   ;-s\right)\right],
\end{split}
\end{equation}
where ${}_0\tilde F_1(;b;z)={}_0 F_1(;b;z)/\Gamma(b)$ is a Regularized ${}_0 F_1$ Hypergeometric Function \citep[chapter 16]{wolffunc,DLMF}. Specifically, 
\[
{}_0\tilde F_1(;b;z)=\sum_{k=0}^\infty \frac{z^k}{\Gamma(b+k)\,k!}.
\]

The evaluation of the integrals (\ref{A1}) and (\ref{A2}) for the $A$-coefficients presents numerical difficulties, as $a_1$ and $a_2$ are highly oscillatory. This is illustrated by the asymptotic behaviour as $s\to\infty$ ($x\to-\infty$) in $-\pi<\arg s \leq\pi$ of the integrand $a_2(s)$ in $A_2$ for the case (\ref{RHU}),
\begin{multline}
a_2(s)\sim\frac{e^{3i\pi/4}\, e^{2 i \sqrt{s} \sec\theta} s^{-i \kappa  \tan \theta }}{\pi \sec ^{5/2}\!\theta}\\
   \biggl[ \left(
   s^{-1/2}+\mathcal{O}\left(s^{-1}\right)\right)\cos \left(2 \sqrt{s}-\pi  \left(\kappa
   +\fract{1}{4}\right)\right)\\
   +\mathcal{O}\left(s^{-1}\right)\sin
   \left(2 \sqrt{s}-\pi  \left(\kappa
   +\fract{1}{4}\right)\right)
  \biggr].        \label{a2}
\end{multline}
Note that 
$
\int s^{-1/2} \exp(2 i \alpha\sqrt{s})\cos2\sqrt{s}\,ds=i(1-\alpha^2)^{-1} \exp(2 i \alpha\sqrt{s})\,[\alpha\cos(2\sqrt{s}-b)-i\sin(2\sqrt{s}-b)]
$
for $\alpha\ne\pm1$, but $\int s^{-1/2} \exp(2 i \sqrt{s})\cos2\sqrt{s}\,ds=\left(i \sqrt{s}-\frac{1}{4} e^{4 i \sqrt{s}}\right) \sin b+\left(\sqrt{s}-\frac{1}{4}i e^{4 i \sqrt{s}}\right) \cos b$.
This indicates that $|A_2(x)|$ diverges as $s\to\infty$ in the resonant case $\sec\theta=\pm1$, but that it converges otherwise, even though $A_2(x)$ itself is oscillatory with finite amplitude. The oscillatory behaviour can be suppressed in $A_2(\infty)$ by integrating along a straight line path $0<|s|<\infty$, $0<\arg s\leq\pi$ rather than directly along the positive real $s$-axis. Doing so renders the numerical evaluation fast and accurate. Similarly, $|A_1(-\infty)|$ is best evaluated using a radius in the lower half complex $s$-plane.

Figure \ref{fig:FPM} shows $\mathscr{F}^+$ and $\mathscr{F}^-$ as functions of $\theta$ and $\kappa$. Several features are immediately apparent:
\begin{enumerate}
\item $\mathscr{F}^+$ and $\mathscr{F}^-$ are the mirror images of each other in $\theta$. This is to be expected, as $A_1(-\infty)$ and $A_2(\infty)$ with opposite signs in $\theta$ are complex conjugates of each other.
\item Both $\mathscr{F}^+$ and $\mathscr{F}^-$ are singular at $\theta=0$, because of the perfect resonant coupling of the Hankel function solutions of the free Alfv\'en waves with the fast wave driving term $R_H$. This is particularly apparent in Equation (\ref{a2}) where the fore-factor $\exp[2i\sqrt{s}]$ (when $\theta=0$) exactly matches with the later cosine term. This renders the $A_1$ and $A_2$ integrals divergent.
\item There is only weak dependence on $\kappa$.
\end{enumerate}

Of course, the conversion coefficients cannot be infinite. Indeed, they cannot exceed 1. Multiplying $\mathscr{F}^+$ and $\mathscr{F}^-$ by the typically very small $\epsilon^2$ to get the actual conversion coefficients $\mathscr{A}^\pm$ restricts the coupling to small $\theta$ (near vertical field), but does not remove the singularity.

There are two reasons for our unphysical results for small $\theta$. First, we have used a perturbation expansion. This clearly breaks down when the coupling is too strong. 

Second, we have assumed an infinite region of coupling. Unlike the ideal MHD coupling by $k_y$ investigated by \citet{CalHan11aa}, which  occurs locally in the neighbourhood of the fast wave reflection point, Hall coupling in our simple model occurs throughout the fast wave's domain. In the Sun and stars though, Hall coupling is associated only with finite regions of very low ionization fraction. As we move deeper into the interior, ionization becomes near-total, and the effect is suppressed. This is illustrated in Figure \ref{fig:FplusP5}, where the contribution to $A_2$ is (arbitrarily) restricted to $x>-5$ (left panel), that is, to the region extending upward from five scale heights beneath the $\kappa=1$ reflection point. Again we see the general preference for small $\theta$, but the amplitudes are reduced by orders of magnitude compared to the full-domain case of Figure \ref{fig:FPM}. The singularity at $\theta=0$ has of course disappeared. Restriction of the contribution to $x>-1$ greatly diminishes the coupling further. The interaction integral $A_2$ is displayed as a function of $x$ or $s$ in Figure \ref{fig:A2} for two field inclinations. It is seen that it is highly oscillatory, but that the oscillatory behaviour is pushed ever deeper as the field becomes more vertical. This limits the conversion in practice for fields of small inclination if the Hall interaction region is of limited extent.

Another feature of Figure \ref{fig:FplusP5} is the shifting of the maximum conversion from vertical field $\theta=0^\circ$ as $\kappa$ increases to $\mathcal{O}(1)$ and above. This is due to the overall wavevector tilting away from the vertical; the maximum is essentially in the field-aligned propagation direction.

Discussion of the application of these insights to the weakly ionized temperature minimum region of the solar atmosphere is deferred till Section \ref{min}.



We turn now to full numerical solution in 2D and 3D.


\section{2D and 3D Numerical Solution}\label{2D3D}
Equations (\ref{parxi}) are now solved using the shooting method as described in \citet{CalHan11aa}, matching on to Frobenius and WKB solutions respectively as $s\to0$ ($x\to\infty$) and $s\to\infty$ ($x\to-\infty$), where $s=e^{-x}$. A fast wave is injected from $s=\infty$, and an outward radiation boundary condition is applied on the Alfv\'en wave at $s=0$. The Hall effect is restricted to a finite region, with 
\begin{equation}
\epsilon(x)=
\begin{cases}
\epsilon_0 r_1(x) & \text{if $x_1-\Delta x<x<x_1$}\\
\epsilon_0 & \text{if $x_1\le x\le x_2$} \\
\epsilon_0 r_2(x) & \text{if $x_2<x<x_2+\Delta x$}\\
0 & \text{otherwise},
\end{cases}
\end{equation}
where $r_1(x)$ and $r_2(x)$ are monotonic quintic polynomials such that $\epsilon(x)$ is twice continuously differentiable throughout. The boundary conditions are applied outside this region. We shall refer to this smoothed top-hat as the ``Hall window'' characterized by parameters $(x_1,x_2,\Delta x)$. In setting $\epsilon$ to zero outside the Hall window, we are assuming the ionization fraction $f$ is insufficiently small there to compensate for $\omega\ll\Omega_i$.

\begin{figure*}[tbhp]
\begin{center}
\includegraphics[width=.82\hsize]{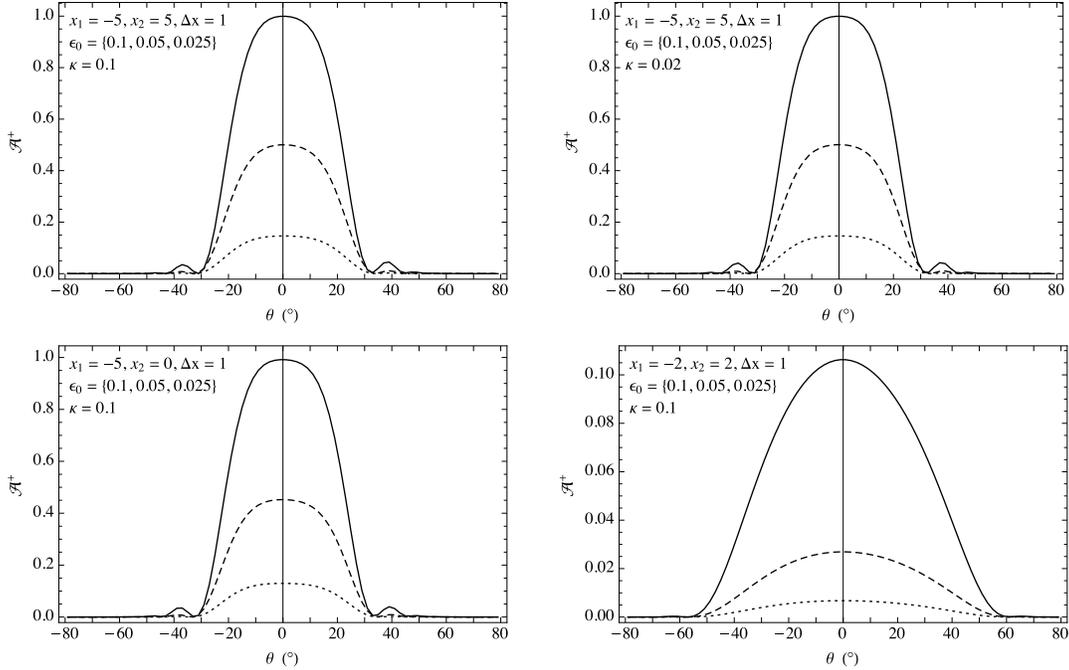}
\caption{Upward fast-to-Alfv\'en conversion coefficients $\mathscr{A}^+$ as a function of magnetic field inclination $\theta$ for a range of parameters in the 2D case $k_y=0$, for which Hall current is the only coupling mechanism. Top left: for Hall window $(-5,5,1)$ with $\kappa=0.1$ and $\epsilon_0=0.1$ (full curve), 0.05 (dashed), and 0.025 (dotted). Top right: the same, but for $\kappa=0.02$. Bottom left: the same as top left, except that the Hall window is restricted to $(-5,0,1)$. Bottom right: the same, but with Hall window (-2,2,1).}
\label{fig:seq2Dtheta}
\end{center}
\end{figure*}

\goodbreak
\subsection{Two Dimensional Cases}
First, consider cases with $k_y=0$, where Hall current is the only mechanism coupling the fast and Alfv\'en waves. Figure \ref{fig:seq2Dtheta} shows the conversion coefficient for a range of parameters as a function of magnetic field inclination $\theta$.

\begin{figure}[bhtp]
\begin{center}
\includegraphics[width=.8\hsize]{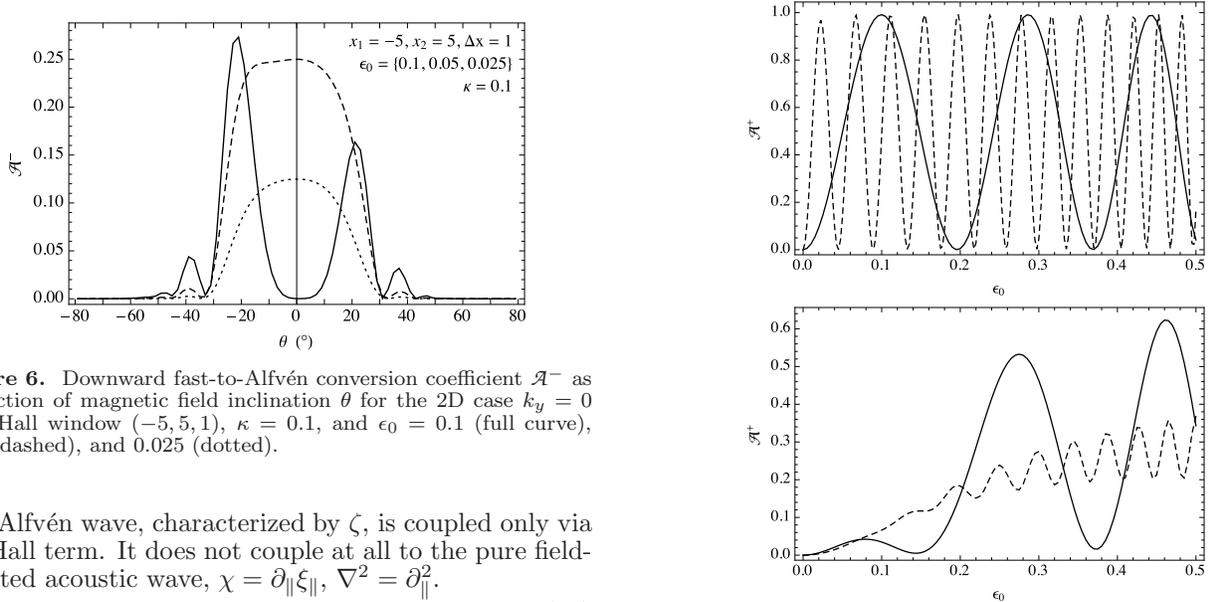}
\caption{Downward fast-to-Alfv\'en conversion coefficient $\mathscr{A}^-$ as a function of magnetic field inclination $\theta$ for the 2D case $k_y=0$ with Hall window $(-5,5,1)$, $\kappa=0.1$, and $\epsilon_0=0.1$ (full curve), 0.05 (dashed), and 0.025 (dotted).}
\label{fig:seq2DthetaMinus}
\end{center}
\end{figure}

The results are entirely consistent with the perturbation results above. Specific points of interest include
\begin{enumerate}
\item For small $\kappa$, conversion is most favoured by vertical or moderately inclined magnetic field, and is very weak for highly inclined field.
\item The dependence on dimensionless wavenumber $\kappa\ll1$ is very weak, since the wave is essentially vertically propagating in any case. 
\item Restriction of the Hall window to $(-5,0,1)$ makes almost no difference compared to $(-5,5,1)$, indicating that the mode conversion is occurring overwhelmingly below $x=0$. For comparison, with $\kappa=0.1$, the fast wave reflection point is $x_\text{ref}=-\ln\kappa^2=\ln100=4.6$. This is very different from $k_y$-mediated conversion, which occurs near $x_\text{ref}$.
\item Restricting the Hall window to $(-2,2,1)$ greatly reduces the mode conversion, indicating that it is chiefly happening below $x=-2$.
\item $\mathscr{A}^+$ is roughly symmetric about $\theta=0^\circ$.
\item Conversion to the downward propagating Alfv\'en wave (Fig.~\ref{fig:seq2DthetaMinus}) is also roughly symmetric for small $\epsilon_0$. However, at large enough $\epsilon_0$, where $\mathscr{A}^+$ approaches 1, it moves to two lobes on either side of the central $\mathscr{A}^+$ peak. Of course, $\mathscr{A}^++\mathscr{A}^-\le1$, so they cannot both be 1 in the same case.
\item Figure \ref{fig:seq2Deps} illustrates the periodic behaviour of $\mathscr{A}^+$ with $\epsilon_0$ at small field inclination that was anticipated in Section \ref{osc}. It is seen that the period increases with increasing height in the atmosphere (increasing Alfv\'en speed), and that it depends on $a$ throughout the window. The periodicity will be discussed again in Section \ref{sec3D}.
\end{enumerate}

Figure \ref{fig:zeta} displays $\zeta=\e_\parallel\vdot\Curl\bxi$ for 2D strongly ($\theta=5^\circ$) and weakly ($\theta=60^\circ$) Hall-coupled cases. The quantity $\zeta$ preferentially selects the Alfv\'en wave. The effect of the lower part of the Hall window is particularly apparent in the strong coupling case with its greatly enhanced $\zeta$-amplitude there.

\subsection{Three Dimensional Cases}  \label{sec3D}
In three dimensions ($k_y\ne0$), the fast and Alfv\'en modes couple, preferentially near the fast wave reflection height, as discussed by \citet{CalHan11aa}. The question now is, to what extent does Hall current modify or add to this conversion?

\begin{figure}[tbhp]
\begin{center}
\includegraphics[width=.8\hsize]{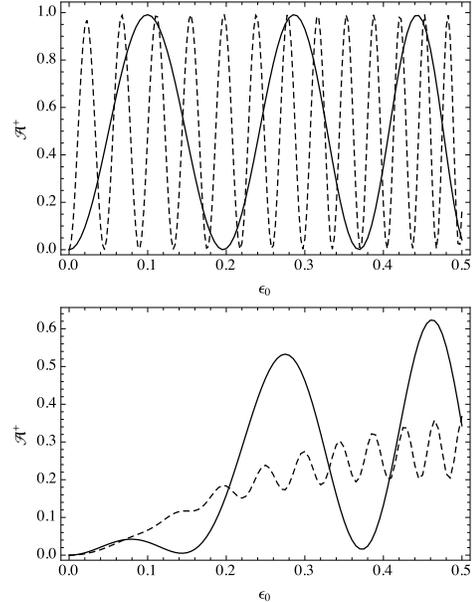}
\caption{Upward Alfv\'en conversion coefficient $\mathscr{A}^+$ as a function of $\epsilon_0$ at $\kappa_z=\kappa=0.1$, $\kappa_y=0$, with $\theta=5^\circ$ (
top frame) and $\theta=30^\circ$ (
bottom frame). The Hall window is alternatively $(-5,5,1)$ (full curve) and $(-8,5,1)$ (dashed curve). 
}
\label{fig:seq2Deps}
\end{center}
\end{figure}

We introduce the orientation angle $\phi$ defined by $\kappa_y=\kappa\sin\phi$ and $\kappa_z=\kappa\cos\phi$. Figure \ref{fig:seq3Dphi} shows how the upward Alfv\'en conversion coefficient $\mathscr{A}^+$ varies with $\phi$ in several cases, compared with the non-Hall case $\epsilon_0=0$. As in 2D, the largest effects are at small $\theta$, reducing to almost nothing at large $\theta$. Specifically,
\begin{enumerate}
\item The Hall parameter $\epsilon_0$ essentially shifts the conversion curve uniformly to the right at small $\theta$. This produces a periodic dependence on $\epsilon$ that was foreshadowed in Section \ref{osc} and seen previously in Figure \ref{fig:seq2Deps}. Effectively, the Hall term adds linearly to the orientation $\phi$, but with a magnitude that increases with depth of the Hall window.
\item At intermediate $\theta$ ($45^\circ$), it enhances the conversion at negative $\phi$ but diminishes it at positive $\phi$.
\item There is practically no effect at $\theta=80^\circ$.
\end{enumerate}

\begin{figure}[tbhp]
\begin{center}
\includegraphics[width=.95\hsize]{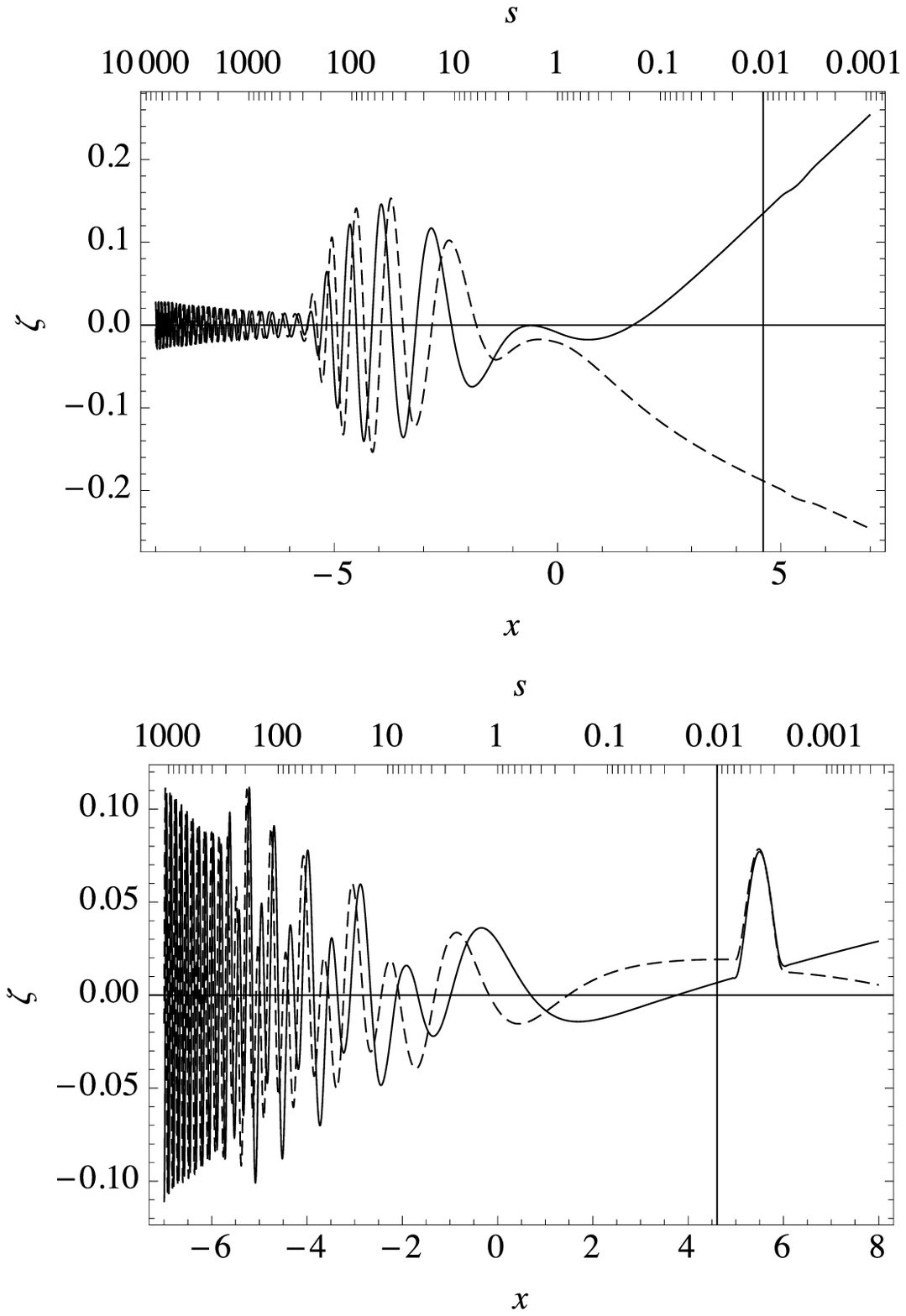}
\caption{The parallel component $\zeta$ of $\Curl\bxi$ in 2D ($k_y=0$) for $\epsilon_0=0.1$, $\kappa=0.1$ with $\theta=5^\circ$ (top) and $60^\circ$ (bottom).  The real and imaginary parts are depicted by the full and dashed curves respectively. The Hall window is $(-5,5,1)$.}
\label{fig:zeta}
\end{center}
\end{figure}

\begin{figure*}[tbhp]
\begin{center}
\includegraphics[width=\hsize]{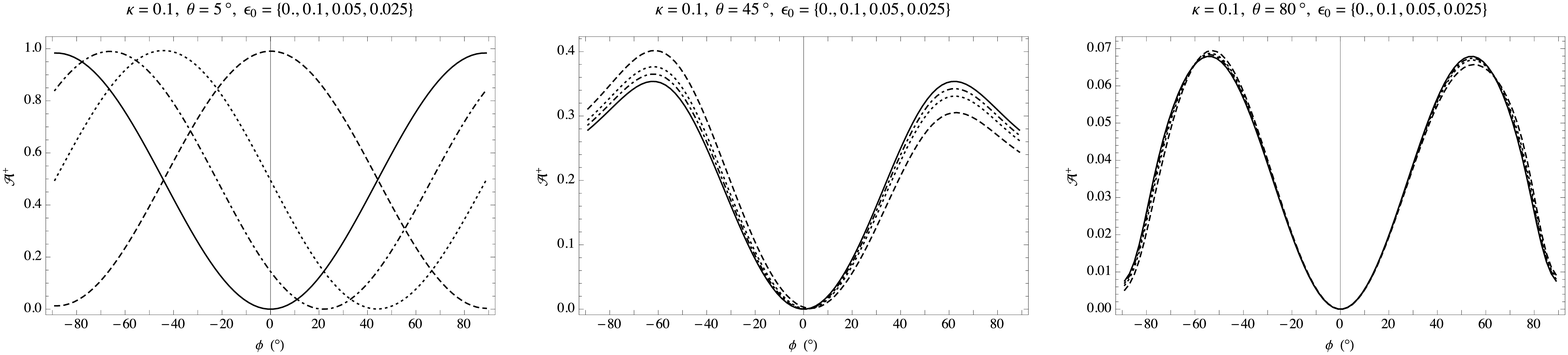}
\caption{Upward Alfv\'en conversion coefficients $\mathscr{A}^+$ against orientation angle $\phi$ for three different magnetic field inclinations: $\theta=5^\circ$ (left frame), $45^\circ$ (centre), and $80^\circ$ (right). In each case, $\kappa=0.1$ and $\epsilon_0=0$ (full curve), 0.1 (dashed), 0.05 (dotted), and 0.025 (chained). The Hall window is $(-5,5,1)$.}
\label{fig:seq3Dphi}
\end{center}
\end{figure*}

\section{Brief Discussion of Warm Plasma Mode Coupling}\label{hot}
The quadratic diminution of fast-to-Alfv\'en mode conversion with increasing magnetic field strength leads us to question the applicability of the cold plasma model $c\ll a$ to the solar photospheric temperature minimum for typical quiet Sun magnetic field strengths. Let us now briefly consider the general warm plasma, $c\ne0$, with particular reference to the high-$\beta$ quiet photosphere. 

Under these circumstances, the Alfv\'en wavelength will be small compared to the density scale height, so for simplicity we neglect gravity, making density $\rho_0$ uniform. Then, from Equation (\ref{waveFull}),
\begin{multline}
\left(\partial_\parallel^2+\frac{\omega^2}{a^2}\right)\bxi  =-\grad_{\!\text p}\chi+\partial_\parallel\grad\xi_\parallel - \frac{c^2}{a^2}\grad\chi\\
+i\,\epsilon\left[\grad\chi\vcross\e_\parallel-\nabla^2\left(\bxi\vcross\e_\parallel\right)\right]
.                               \label{hoteqn}
\end{multline}
Note that the plasma displacement $\bxi$ is no longer restricted to be perpendicular to the magnetic field lines.


Equation (\ref{hoteqn}) is conveniently split into three scalar equations by taking respectively the component parallel to the magnetic field:
\begin{subequations}\label{hotcmpts}
\begin{equation}
\omega^2\xi_\parallel+c^2 \partial_\parallel\chi =0;
\end{equation}
the divergence:
\begin{equation}
\left[\omega^2+(a^2+c^2)\nabla^2\right]\chi-a^2\nabla^2\partial_\parallel\xi_\parallel+i\thth\epsilon\thth a^2\nabla^2\zeta=0;
\end{equation}
and the parallel component of the curl:
\begin{equation}
\left(a^2\partial_\parallel^2+\omega^2\right)\zeta = i\thth\epsilon\thth a^2\left(\partial_\parallel^2\chi-\nabla^2\partial_\parallel\xi_\parallel\right).
\end{equation}
\end{subequations}
The Alfv\'en wave, characterized by $\zeta$, is coupled only via the Hall term. It does not couple at all to the pure field-directed acoustic wave, $\chi=\partial_\parallel\xi_\parallel$, $\nabla^2=\partial_\parallel^2$.

The dispersion relation associated with Equations (\ref{hotcmpts}) is
\begin{multline}
\left(\omega^2-a^2k_\parallel^2\right)\left(\omega^4-(a^2+c^2)\omega^2k^2+a^2c^2k_\parallel^2 k^2\right)\\
=\epsilon^2a^4k_\parallel^2 k^2 (\omega^2-c^2k^2),  \label{hotDR}
\end{multline}
providing a neat generalization of the standard ideal MHD result \citep[Sec.~5.2.3]{GoePoe04aa}, where the right hand side vanishes.

The parallel propagation case $k=k_\parallel$ reduces to
\begin{equation}
\left(\omega^2-c^2k^2\right)
\left( (\omega^2-a^2k^2)^2-\epsilon^2a^4 k^4\right)=0.
\end{equation}
The acoustic mode decouples, leaving exactly Equation (\ref{parDR}), describing Hall-driven precession of the joint magnetically dominated magneto\-acoustic and Alfv\'en modes. For this case at least, the warm plasma will display exactly the same Hall-mediated oscillation of the Alfv\'en and magneto-acoustic modes as the cold plasma, though if $a<c$ it will be the slow rather than the fast wave that is involved. Away from the field-aligned direction, sound speed plays a part, so details will differ. Warm plasma coupling will not be considered further here.

\section{Why the Oscillation?}

The Hall effect typically acts as a catalyst to MHD processes. It does not in itself add energy to a system, or power instabilities. It can however facilitate the transfer of energy from shear flows for example to oscillations or instabilities 
(\citealt{War99aa};  \citealt{PanWar12aa}; \citealt{RudKitHol13aa}, Sections 2.7, 5.5, and 6.6). The Hall effect is also known to facilitate reconnection in an indirect way {\color{black}\citep[see][and references therein]{ShaDraRog01aa,Mal08aa}.}

The process we have identified here is physically similar: the Hall terms transfer energy from the oscillatory shear flow of an Alfv\'en wave to the compressive fast wave, and \emph{vice versa}. This is particularly apparent in Equation (\ref{chizeta}), where we see a Hall-mediated transfer of energy from the shear flow to the compressional oscillation $\chi$ described by the shear-sensitive term $\epsilon\nabla^2\zeta$ on the right hand side, and a corresponding flow back from compression to Alfv\'enic shear through $\epsilon\,\partial_\parallel^2\chi$.


 In a way, the conversion process we identify here is similar to fast-to-Alfv\'en conversion in ideal MHD mediated by non-zero $k_y$. In both cases we need a situation where the phase velocities of both waves are nearly aligned and of the same magnitude so the energy can be transferred between them. In the $k_y$-mediated case this is achieved via variations of the stratification. In the Hall-mediated case considered here, this is achieved via aligning both waves close to the magnetic field, so a small addition of an $\epsilon$-proportional contribution facilitates the transfer of the energy from fast to Alfv\'en and \emph{vice versa}. Physically, in the $k_y$-mediated case the fast and Alfv\'en waves both quickly leave the conversion area, each of them following their own distinct path. However, in the Hall-mediated case, in the region of maximum conversion, the waves keep propagating long distances nearly aligned with the field since the conversion happens far away from the fast wave reflection point, so nothing prohibits them from transferring their energies back and forth via precession as they travel through the Hall window.


\section{Application to the Sun's Temperature Minimum, Reconnection, and Star Forming Regions}\label{min}
\subsection{\color{black}Temperature Minimum}
Taking ion number density data from the quiet Sun Model C of \citet[Tables 12 and 17--24]{VerAvrLoe81aa}, based on a mixture of neutral and singly ionized H, He, C, Mg, Al, Si, and Fe, the mean atomic weight of ions at the temperature minimum is $m_\text{i}=38.7$ amu, the total ion number density is $n_\text{i}=2.29\times10^{17}$ $\rm m^{-3}$, the mean ion gyrofrequency is $\Omega_i=2.47\times10^6B_0$ rad $\rm s^{-1}$, where $B_0$ is measured in Tesla, and the ionization fraction $m_\text{i}n_\text{i}/\rho$ is $f=3\times10^{-3}$. The electron number density $n_\text{e}=n_\text{i}$ is slightly below the value $2.495\times10^{17}$ $\rm m^{-3}$ listed in Table 12, presumably because of the presence of other minor ions, but is close enough for our purposes. For a 5 mHz wave, this gives $\epsilon=4.2\times10^{-6}B_0^{-1}$. The Alfv\'en speed is $a=400B_0$ $\rm km\,s^{-1}$. For a Hall window of width $L$, measured in km, the oscillation number is therefore $N_\text{osc}\approx5\times10^{-11}B_0^{-2}L$. At $B_0=10^{-4}$ T (1 G) with $L\approx 600$ km, we have oscillation number $N_\text{osc}\approx3$, but this diminishes quadratically with increasing field strength. It is already negligible for 10 G.

One way of enhancing Hall conversion is for the magnetic field to be highly inclined, so that the effective $L$ is greatly increased, and the wave propagation direction angled to match. This requires $\kappa\gg1$, but we do not expect much power at such wavenumbers in the $p$-mode spectrum. However, other locally excited waves may exhibit such behaviour.

Ionization fractions are lower in sunspot umbrae \citep{MalAvrCar86aa}, but their much greater magnetic field strengths more than compensate, resulting in even smaller oscillation numbers.

Although Hall-mediated conversion \textcolor{black}{of low frequency waves} is only effective for the very weak fields of the quiet Sun, the overall amount of converted energy can still be significant since the quiet Sun occupies most of solar volume.  Indeed, recent high-resolution measurements using Hanle and Zeeman effects reveal that the quiet Sun is full of weak magnetic fields of magnitude below 10--100 G \citep{TruShcAse04aa,SanMar11aa}.  Therefore, the process identified here can contribute to chromospheric and coronal heating by facilitating the propagation of Alfv\'en waves into the upper atmosphere in quiet solar areas.

\subsection{\color{black}Reconnection Events}
{\color{black}
The Hall effect has already been identified in simulations as crucial to the process of fast reconnection in collisionless plasmas by allowing inflow close to the Alfv\'en speed \citep{ShaDraRog01aa,RogDenDra01aa,BirDraSha01aa}. If, as commonly believed, small scale reconnection is ubiquitous through the solar atmosphere, spanning both collisional and collisionless regimes, fast/Alfv\'en oscillation may play a significant role in light of the radically higher frequencies involved.

Escaping waves originating from these events may also be affected. It was suggested by \citet{AxfMcK92aa} that high frequency waves with periods of order or less than one second can be released in `microflare' reconnections. By increasing $\epsilon$ by between 2 and 3 orders of magnitude, all other things being equal, reconnection waves would indeed be most susceptible to Hall-mediated conversion.}

\subsection{\color{black}Star Forming Regions}
Turning now to star forming regions, we take some typical values from the simulations of \citet{WurPriBat15aa}, who find that Hall current aids collapse and disc formation if the magnetic field is anti-aligned with the rotation axis. Specifically, we set $B_0=4\times10^{-8}$ T, $\rho_0=7.5\times10^{-15}$ $\rm kg\,m^{-3}$, mean ion mass 24.3 amu, and initial cloud size $L=0.013$ parsec. The Alfv\'en speed is then about 400 $\rm m\,s^{-1}$, and the ion gyrofrequency is 0.16 $\rm s^{-1}$. Crucially, the ionization fraction is $f=10^{-10}$ or lower. We take as the dynamical timescale the free-fall time of $T=2.4\times10^4$ yr and consider an Alfv\'en period of that order: $\omega=2\pi/T=8.3\times10^{-12}$ $\rm s^{-1}$. These give $\epsilon\sim0.5$, and $N_\text{osc}\sim0.7$.

These numbers are all very rough, and the result could vary by large amounts either way, but it is interesting that the oscillation number is at least plausibly of $\mathcal{O}(1)$. We might hypothesise that the process could even contribute to star formation by taking energy from incompressive Alfv\'en waves and converting it to compressive magneto\-acoustic waves. Any increase in density could help precipitate gravitational collapse.

\section{Conclusions}\label{conc}
The main purpose of this paper is to explore the basic principles of Hall coupling between magneto\-acoustic and Alfv\'en waves in weakly ionized plasmas. To that end, we have focussed on the simplest scenario, a cold plasma, supporting only fast and Alfv\'en waves.

The potential coupling of these two wave types by the ``precessional effect'' of Hall current in a uniform plasma with field-aligned wavevector has been confirmed for stratified and unstratified plasmas with arbitrary wave direction. The basic findings are:
\begin{enumerate}
\item Hall-mediated conversion of near-vertically propagating seismic waves in the Sun's surface layers primarily occurs for small to moderate magnetic field inclinations from the vertical and at sufficient depth that the wavevector is closely aligned with the field, unlike the $k_y$-mediated conversion.
\item The process occurs throughout regions satisfying these criteria, with the total conversion coefficient determined by an integral across them ($A_2$ in the 2D case, or the oscillation number $N_\text{osc}$ in weakly stratified atmospheres). It is therefore crucial to accurately identify the regions that support Hall current (the ``Hall window'').
\item The strength of the effect is determined by the Hall parameter $\epsilon=\omega/(f\Omega_\text{i})$, where $\omega$ is the wave frequency (typically a few mHz), $\Omega_\text{i}$ is the ion gyrofrequency (several MHz), and $f$ is the ionization fraction. The upward and downward mode conversion coefficients $\mathscr{A}^+$ and $\mathscr{A}^-$ scale quadratically with $\epsilon$, except at larger values where they saturate at 1.
\item The previously identified ``precession of the polarization'' of field-aligned Alfv\'en waves due to Hall current is in fact not simply the precession of a single mode. It is instead an oscillation between fast and Alfv\'en modes. This is not just a semantic point: fast and Alfv\'en waves have distinct dispersion relations, with the former subject to reflection and the latter not. This has major implications for determining the source of transverse waves observed and inferred in the solar corona.
\item In principle, a wave propagating nearly parallel to the magnetic field oscillates between fast and Alfv\'en states approximately $N_\text{osc}=\int_{x_1}^{x_2}\epsilon\thth\omega/2\pi a\, dx$ times in passing through a Hall window $x_1<x<x_2$. The final state emerging at $x_2$ therefore depends on how close this is to being an integer or half-integer. 
\item The oscillation number $N_\text{osc}=\int_{x_1}^{x_2}\epsilon\thth\omega/2\pi a\, dx$ depends inverse-quadratically on the magnetic field strength. One factor of $B_0^{-1}$ resides in $\epsilon$, through the ion gyrofrequency, which describes a rate of oscillation \emph{per unit time}. However, another results from the inverse dependence on the Alfv\'en speed of the oscillation number \emph{per unit distance}, that is more relevant to the issue of the conversion coefficient of a fixed Hall-effective layer.
\item For ionization fractions of a few times $10^{-3}$, characteristic of the quiet Sun temperature minimum, significant Hall-mediated conversion of low frequency waves is apparently restricted to regions of weak magnetic field, no more than a few Gauss. \textcolor{black}{However, higher frequency waves generated in reconnection and other violent small scale events should exhibit Hall-mediated oscillation between modes even at significantly higher field strengths.}
\item Sunspot (kilogauss) strength magnetic fields should not exhibit significant Hall coupling at mHz frequencies because of their high (MHz) ion gyrofrequencies, which make $\epsilon$ too small, exacerbated by their high Alfv\'en speeds, further reducing the oscillation number.
\item In reality, we probably need magnetic fields of over 100 G for the cold plasma approximation to be even marginally viable in the Sun's temperature minimum region, so our simple cold plasma model may not be applicable to that scenario. Nevertheless, we have elucidated the nature of the coupling, and shown it not to be important at much greater field strengths. We have also demonstrated in Section \ref{hot} that Hall-mediated oscillation in the field-aligned case is independent of sound speed, so will continue to operate whatever the plasma $\beta$. The full warm plasma coupling will be examined in realistic atmospheres in Paper II.
\item Characteristic parameters for star forming gas clouds suggest that mode conversion/oscillation may operate effectively there.
\end{enumerate}

Irrespective of the level of applicability in specific regions of the Sun, stars, and interstellar medium, the process of Hall-mediated oscillation between Alfv\'en and magneto\-acoustic waves is a fundamental \textcolor{black}{and interesting} aspect of Hall-MHD. \textcolor{black}{It is also a good benchmark mechanism for Hall-MHD codes.}

%
\appendix
\section{Derivation of Cold Plasma Wave Equation}\label{app:der}
The linearized momentum, induction, and current density equations are respectively 
\begin{gather}
-\rho_0\omega^2\bxi = \bj_1\vcross\B_0 +\F 
\label{mmntm}\\
-i\,\omega\thth\B_1 =-i\,\omega\Curl(\bxi\vcross\B_0)- \Curl\frac{\bj_1\vcross\B_0}{e\thth n_\text{e}}  \label{induct}\\
\bj_1=\frac{1}{\mu_0}\Curl\B_1  \label{curr}
\end{gather}
where the 0 subscript denotes the background value, and 1 the Eulerian perturbation, and an $\exp[-i\,\omega\,t]$ time dependence is assumed. The thermal and gravitational terms are included at this stage and collected as $\F=\rho_1\g-\grad p_1$, where $\g=-g\e_x$ is the gravitational acceleration, $\rho_1=-\rho_0(\chi-\xi_x/h)$ and $p_1=-\rho_0(c^2\chi- g\xi_x)$ is the gas pressure perturbation. We eliminate $\bj_1\vcross\B_0$ between Equations (\ref{mmntm}) and (\ref{induct}), and substitute the resulting expression for $\B_1$ into Equation (\ref{curr}). This expression for $\bj_1$ is then substituted back into Equation (\ref{mmntm}), leaving an equation for $\bxi$ only:
\begin{equation}
\frac{\omega^2}{a^2}\bxi=-\left[\Curl\Curl(\bxi\vcross\hat\B_0)\right]\!\vcross\hat\B_0
+i\left(\curl\curl\epsilon\thth\bxi\right)\!\vcross\hat\B_0
+\fbox{$\displaystyle i\thth\epsilon f\,\frac{\left[\curl\curl\rho_1\g\right]\vcross\hat\B_0}{\rho_0\omega^2}-\frac{\F}{\rho_0a^2}$}.  \label{waveFull}
\end{equation}
The terms in the box vanish in the limit $c^2/a^2\to0$ for waves of interest, $\omega\sim a k$, noting that $gh=\mathcal{O}(c^2)$ by hydrostatic equilibrium. The first boxed term is very small in any case due to the factor $f$.

The use of various vector identities reduces the remaining terms to Equation (\ref{basiceqn}).

\acknowledgments
This work is partially supported by the Spanish Ministry of Science through projects AYA2010-18029, AYA2011-24808, and AYA2014-55078-P. This work contributes to the deliverables identified in FP7 European Research Council grant agreement 277829, ``Magnetic connectivity through the Solar Partially Ionized Atmosphere''.

\vspace{20pt}

\bibliographystyle{apj}        
\bibliography{fred}

\end{document}